\documentclass[conference]{IEEEtran}
\IEEEoverridecommandlockouts
\usepackage{cite}
\usepackage{amsmath,amssymb,amsfonts}
\usepackage{algorithmic}
\usepackage{graphicx}
\usepackage{textcomp}
\usepackage{amsthm,amsmath,amssymb}
\usepackage{stmaryrd,txfonts}
\usepackage{mathrsfs}
\usepackage{algorithm}
\usepackage{threeparttable}
\newtheorem{prop}{Proposition}
\begin{document}

\title{An Optimization Model for Offline Scheduling Policy of Low-density Parity-check Codes}
\author{\IEEEauthorblockN{Dongxu Chang \IEEEauthorrefmark{2},
                Zhiming Ma \IEEEauthorrefmark{3}, 
                     Guanghui Wang \IEEEauthorrefmark{1}\IEEEauthorrefmark{2},
                     Guiying Yan
                     \IEEEauthorrefmark{3},
                     Dawei Yin \IEEEauthorrefmark{1},}
    \IEEEauthorblockA{\IEEEauthorrefmark{1}%
                     School of Mathematics, Shandong University
                     \IEEEauthorrefmark{2}%
                     Data Science Institute, Shandong University\\
                     \IEEEauthorrefmark{3}%
                     Academy of Mathematics and Systems Science, University of Chinese Academy of Sciences}
                     
    Email: dongxuchang@mail.sdu.edu.cn, mazm@amt.ac.cn, ghwang@sdu.edu.cn,\\ yangy@amss.ac.cn, daweiyin@mail.sdu.edu.cn
           }

\maketitle

\begin{abstract}
In this study, an optimization model for offline scheduling policy of low-density parity-check (LDPC) codes is proposed to improve the decoding efficiency of the belief propagation (BP). The optimization model uses the number of messages passed (NMP) as a metric to evaluate complexity, and two metrics, average entropy (AE), and gap to maximum $\textit{a posteriori}$ (GAP), to evaluate BP decoding performance. Based on this model, an algorithm is proposed to optimize the scheduling sequence for reduced decoding complexity and superior performance compared to layered BP. We validated the proposed algorithm on LDPC codes constructed following 5G New Radio, which resulted in a reduction of decoding complexity of more than 20$\%$ compared to LBP.
\end{abstract}

\begin{IEEEkeywords}
Low-density parity-check code, scheduling policy, decoding efficiency
\end{IEEEkeywords}

\section{Introduction}
\label{sec:introduction}
Low-density parity-check (LDPC) code\cite{gallager1962low} has been widely used in communication fields, such as 5G New Radio (NR)\cite{3gpp20185g}, IEEE 802.11 (WiFi)\cite{ieee2007ieee}. It was demonstrated by D. MacKay and M. Neal in 1996\cite{mackay1997near} that LDPC codes can approach the Shannon limit. The belief propagation (BP) decoding algorithm\cite{mackay1999good} of LDPC code is most commonly used due to its low complexity and good performance in practice.

The conventional BP algorithm employs a flooding scheduling strategy to conduct message-passing between variable nodes and check nodes. This strategy updates all variable-to-check (V2C) messages or check-to-variable (C2V) messages simultaneously in one iteration, achieving a high degree of parallelism. However, the flooding BP algorithm is known for its slow convergence speed. This is because the latest information available in the current iteration can only be utilized in the subsequent iteration. The delayed information may hinder the benefits of information transmission, thus affecting the algorithm's convergence rate\cite{hocevar2004reduced}. 

In 2004, D. Hocevar introduced the concept of layered belief propagation (LBP) decoding\cite{hocevar2004reduced}, which updates information sequentially. This decoding strategy can accelerate the decoding process as it ensures that the most recent information is disseminated. In comparison to the flooding scheduling strategy, LBP can reduce decoding complexity by nearly 50$\%$ while achieving the same performance.

Despite its benefits, LBP does not have a specific design for the decoding sequence order. To further reduce decoding complexity, scheduling the order of decoding sequences is a key direction for improving LBP. In the following, we define the order of transmitting information among nodes or edges in decoding as a scheduling sequence. In \cite{casado2007informed}\cite{gong2011effective}\cite{liu2014variable}, several online (dynamic) scheduling policies have been proposed. These policies determine the scheduling sequences based on the latest information from the previous updates, meaning the decoding order is established during the decoding process. However, additional calculations are required to determine the choice of the next part of the scheduling sequences. In some cases, these calculations can be even more complex than the decoding process itself, which may contradict the original intention of scheduling optimization and result in high latency.

In contrast to online scheduling policies, offline scheduling policies are fixed for a given decoder and achieve high efficiency in statistical significance without adding extra complexity of determining the scheduling sequence to the decoding process. Consequently, offline scheduling policies can better align with practical application scenarios than online scheduling policies. Several relevant studies have been conducted on this topic, including \cite{wang2020two}\cite{frenzel2019static}\cite{kim2012serial}\cite{jang2022design}. In \cite{wang2020two}, the Least-punctured and highest-degree (LPHD) algorithm is proposed, in which check nodes with higher degrees and fewer punctured variable nodes are updated first. In \cite{frenzel2019static}, Low-degree (LD) schedule gives priority to updating check nodes with a lower degree. In \cite{kim2012serial}, a scheduling policy is proposed that considers the characteristics of the parity-check matrix of LDPC to ensure that updated messages are propagated to other nodes as much as possible. These algorithms design scheduling sequences by analyzing different aspects of the parameters of the decoding factor graph but leave little room for further improvement. In \cite{jang2022design}, a nested scheduling policy is designed in a greedy manner based on layered reciprocal channel approximation based density evolution. Nevertheless, an offline scheduling policy that captures the scheduling operation more analytically through a better optimization algorithm is required.

In this study, we model the efficiency of scheduling policies using a complexity-performance paradigm. The complexity is measured by the number of messages passed (NMP) \cite{yin2022message}, which counts the message passing on directed edges to measure complexity. For the performance evaluation of scheduling policies, we consider two different metrics. The first one is average entropy (AE), which measures the uncertainty of variable nodes after each message passed. It calculates the LLR distribution of the messages sent by the variable nodes using density evolution \cite{richardson2001capacity} and then averages the entropy of the variable node information. The second metric is gap to maximum $\textit{a posteriori}$ (GAP) \cite{yin2022message}. Based on the Bethe free energy, GAP measures the statistical distance between the distribution obtained by message passing and the true posterior distribution. By optimizing the value of AE or GAP of the scheduling sequences, we can design an algorithm to optimize the scheduling sequences. A simple greedy scheduling policy based on GAP has already been proposed in \cite{yin2022messagearxiv}. It can achieve a $10\%$ speed increase compared to LBP when the number of iterations is small. However, the performance of this greedy scheduling policy can significantly degrade when the number of iterations increases, as it only considers the influence of the scheduling sequence on the current decoding process, neglecting the longer-term effects on the decoding. 

In this study, we present a novel algorithm, called the Successive-Searching BP (SSBP), which is based on the local search algorithm \cite{aarts2003local} and the complexity-performance paradigm. Local search algorithms have a long history of success in combinatorial optimization problems, and the SSBP algorithm can effectively search for scheduling sequences that lead to fast declines in AE or GAP throughout the decoding process. We also analyze the average-case computational complexity of the SSBP algorithm. Simulation results demonstrate that the scheduling sequences generated by SSBP exhibit significantly lower decoding complexity and better error-rate performance than those generated by other scheduling policies.

The rest of the paper is organized as follows: Section \uppercase\expandafter{\romannumeral2} provides an overview of relevant LDPC decoding algorithms and introduces the metrics used in the complexity-performance paradigm. The proposed SSBP algorithm is described and analyzed in Section \uppercase\expandafter{\romannumeral3}. The simulation results of our algorithm are presented in Section \uppercase\expandafter{\romannumeral4}. We conclude the paper in Section \uppercase\expandafter{\romannumeral5}.

\section{Preliminaries}

This section will begin by providing a brief introduction to some basic definitions of LDPC codes, followed by an overview of several related LDPC decoding algorithms. We will then introduce our metric for measuring decoding complexity, along with two metrics used to measure decoding performance.

\subsection{Basic definitions of LDPC codes}
An LDPC code is a linear block code with a sparse parity check matrix \cite{gallager1962low}, which can be represented by a bipartite graph with variable nodes and check nodes. The parity check matrix of a binary $(N, K)$ LDPC code $\mathcal{C}$ of rate $R=K/N$ is a 0-1 matrix of size $(N-K)\times N$, in which each column of the matrix corresponds to a variable node and each row corresponds to a check node. The entries of the matrix determine whether the corresponding variable nodes and check nodes are connected in the associated bipartite graph. 

\subsection{Relative LDPC decoding schemes}
\subsubsection{Flooding BP}
The message-passing process of BP decoding consists of V2C message updates and C2V message updates. The message update rule of V2C is
\begin{equation}
    m_{i \rightarrow \alpha}^{(l)}=m_{0}+\sum_{h \in N(i) \backslash \alpha} m_{h \rightarrow i}^{(l-1)},
\end{equation}
and the message update rule of C2V is 
\begin{equation}
    m_{\alpha \rightarrow i}^{(l)}=2 \tanh ^{-1}\left(\prod_{j \in N(\alpha) \backslash i} \tanh \left(m_{j \rightarrow \alpha}^{(l-1)} / 2\right)\right)
\end{equation}
where $m_0$ is the channel message in LLR form, $l$ is the number of iterations, $N(v)$ represents the nodes connected directly to node $v$, $m_{i \rightarrow \alpha}^{(l)}$ means the message from variable node $i$ to check node $\alpha$ in iteration $l$ and $m_{\alpha \rightarrow i}^{(l)}$ means the message from check node $\alpha$ to variable node $i$ in iteration $l$. The initial message $m_{i \rightarrow \alpha}^{(0)}$ and $m_{\alpha \rightarrow i}^{(0)}$ is 0.

In flooding BP, during a single iteration, all check nodes' information is first updated using (1). After that, (2) updates all the variable nodes. The above process will repeat until all the parity-check equations are satisfied or the maximum number of iterations is reached. As previously discussed, the new message obtained by the nodes can only be used in the next iteration, which can impair both the convergence speed and the error-rate performance.

\subsubsection{Layered BP (LBP)\cite{hocevar2004reduced}}
 Different from flooding BP, layered BP updates the C2V message right after one layer performs V2C updates. The V2C update rule of the above process is
\begin{equation}
    m_{i \rightarrow \alpha}^{(l)} = m_{i}^{(l-1)}-m_{\alpha \rightarrow  i}^{(l-1)},
\end{equation}
where $m_{i}$ is the posterior probability of the variable node $i$. At the end of each layer, the posterior probability value should be updated as
\begin{equation}
    m_i^{(l)}=m_{i\rightarrow\alpha}^{(l)}+m_{\alpha\rightarrow i}^{(l)}.
\end{equation}
The conventional layered BP has no specific design for the scheduling sequences, which can lead to imperfect performance since different schedule sequences can affect the result of the message-passing process to a large extent.

\subsection{The metric to measure the decoding complexity}
\subsubsection*{Number of messages passed (NMP)\cite{yin2022message}}
NMP is a metric used to measure the complexity of the decoding process by defining the V2C or C2V message passing on a directed edge as a unit of complexity. The complexity is determined by the number of messages passed during decoding, and the incremental value of NMP after decoding on a check node with degree $d_c$ is $2d_c$ due to the requirement of $d_c$ input messages and $d_c$ output messages. The average value of NMP can be used to compare the decoding complexity under different scheduling policies through simulation experiments.

\subsection{Two metrics to measure the decoding performance}

\subsubsection{Average entropy (AE)}
The BP process can be analyzed by density evolution \cite{richardson2001capacity}. In the following, the concepts and notations used in \cite{richardson2008modern} will be employed. Assume that $\boldsymbol{x}=\left\{x_1,\cdots,x_n\right\}\in\left\{0,1\right\}^n$ are transmitted through binary memoryless symmetric channel (BMS) $W$, and $\boldsymbol{y}=\left\{y_1,\cdots,y_n\right\}$ are received signals. Let
\begin{equation}
    L(y_i)=\ln{\frac{p(x_i=0|y_i)}{p(x_i=1|y_i)}}
\end{equation}
denote the corresponding LLR of $y_i$, and $c_i$ be the density of $L(y_i)$ conditioned on $x_i=0$. We call $c_i$ the L-density of $y_i$. The L-density of $m_{\alpha\rightarrow i}$ and $m_{i\rightarrow \alpha}$ are denoted by $c_{(\alpha,i)}$ and $c_{(i,\alpha)}$, respectively. The entropy function of L-density is the linear function defined by
\begin{equation}
    H(c)\triangleq -\int_{-\infty}^{+\infty}c(y)\log_2(1+e^{-y})dy.
\end{equation}

Denote the convolution operations on the variable node and check node by two binary operators $\circledast$ and $\boxast$, respectively. For L-density $c_1$, $c_2$, and any Borel set $E\in \overline{\mathbb{R}}$, define
\begin{equation}
    (c_1 \circledast c_2)(E)\triangleq \int_{\overline{\mathbb{R}}} c_1(E-\alpha)c_2(d\alpha),
\end{equation}
\begin{equation}
    (c_1 \boxast c_2)(E)\triangleq \int_{\overline{\mathbb{R}}} c_1\left(2\tanh^{-1}\left(\frac{\tanh{\frac{E}{2}}}{\tanh{\frac{\alpha}{2}}}\right)\right)c_2(d\alpha),
\end{equation}
where $\int_{\overline{\mathbb{R}}}f(\alpha)c(d\alpha)$ is the Lebesgue integral with respect to probability measure $c$ on extended real numbers $\overline{\mathbb{R}}$. Using density evolution, we can get the L-density $c_i$, $c_{(\alpha,i)}$ and $c_{(i,\alpha)}$ as 
\begin{equation}
    c_{(\alpha,i)}^{(0)}=c_{(i,\alpha)}^{(0)} = \Delta_0,
\end{equation}
\begin{equation}
    c_{(\alpha,i)}^{(t)}=\boxast_{j\in N(\alpha)\backslash i}c_{(j,\alpha)}^{(t-1)},
\end{equation}
\begin{equation}
    c_{(i,\alpha)}^{(t)} =c_i\circledast\left(\circledast_{h\in N(i)\backslash\alpha}c_{(h,i)}^{(t-1)}\right),
\end{equation}
where $\Delta_0$ is the greatest element defined by a partial order on the space of symmetric probability measures.

Let $\mathcal{X}$ be a convex subset of symmetric probability measures, $\mathbb{N}$ be the set of natural numbers, and $\mathbb{R}$ be the set of real numbers. The AE function, $E_{\beta}:\mathcal{X}\times\mathbb{N}\rightarrow\mathbb{R}$, of the channel messages with the L-density $\boldsymbol{c}=(c_1,\cdots,c_N)$ and number of messages passed $t$ under scheduling sequence $\boldsymbol{\beta}$ is 
\begin{equation}
    AE_{\boldsymbol{\beta}}\left(\boldsymbol{c},t\right)\triangleq\frac{1}{N}\left\{\sum_i H\left(c_i\circledast\left(\circledast_{\alpha\in N(i)}c_{(\alpha,i)}^{(t)}\right)\right)\right\}.
\end{equation}

%AE can reflect the remaining uncertainty of the current decoding system and can be regarded as an upper bound of the bit error rate. Let $D_i(y_1,\cdots,y_N)$ be the maximum a posteriori (MAP) decoder's estimation of $x_i$ given $y_1,\cdots,y_N$. Using the Kovalevskij inequality \cite{ho2010interplay}, we can derive:
%\begin{equation}
%\begin{split}
%    2P(x_i\neq D_i(y_1,\cdots,y_N)) &\leq H(x_i|y_1,\cdots,y_N)\\ &\leq H\left(c_i\circledast\left(\circledast_{\alpha\in N(i)}c_{(\alpha,i)}^{(t)}\right)\right).
%\end{split}
%\end{equation}
%As the BP decoder approximates the MAP decoder, the error-rate of the BP decoder can also be linked to AE. 

\subsubsection{Gap to maximum a posteriori (GAP)\cite{yin2022message}}
GAP is another metric to measure decoding performance. Based on the Bethe free energy, GAP measures the statistical distance between the distribution obtained by message passing and the true posterior distribution.

The GAP function, $GAP_{\beta}:\mathcal{X}\times\mathbb{N}\rightarrow\mathbb{R}$, of the channel messages with the L-density $\boldsymbol{c}=(c_1,\cdots,c_N)$ and number of messages passed $t$ under scheduling sequence $\boldsymbol{\beta}$ is 
\begin{equation}
    \begin{split}
        GAP_{\boldsymbol{\beta}}\left(\boldsymbol{c},t\right)
        &\triangleq-\frac{1}{N}\bigg\{\sum_i H\left(c_i\circledast\left(\circledast_{\alpha\in N(i)}c_{(\alpha,i)}^{(t)}\right)\right)\\
        &+\sum_\alpha \sum_{i\in N(\alpha)}H\left( c_{(i,\alpha)}^{(t)}\right)-\sum_{\alpha}H\left(\boxast_{i\in N(\alpha)}c_{(i,\alpha)}^{(t)}\right)\\
        &
         -\sum_{(i,\alpha)}H\left(c_{(i,\alpha)}^{(t)}\circledast c_{(\alpha,i)}^{(t)} \right) \bigg\}.
    \end{split}
\end{equation}

\section{Scheduling Strategies}
\subsection{SSBP$\textendash$new scheduling strategy}
As AE and GAP can both reflect the performance of the decoding process, we can use $AE_{\boldsymbol{\beta}}(\boldsymbol{c},t)$ (or $GAP_{\boldsymbol{\beta}}(\boldsymbol{c},t)$) to evaluate the decoding system under scheduling sequence $\boldsymbol{\beta}$. A scheduling sequence with a rapid decline of the corresponding $AE_{\boldsymbol{\beta}}(\boldsymbol{c},t)$ (or $GAP_{\boldsymbol{\beta}}(\boldsymbol{c},t)$) throughout the decoding process is desirable.

However, the influence of the decline in the AE function (or the GAP function) on the decoding complexity varies depending on the decoding period. In practice, achieving high throughput requires a decoder that can decode successfully with few iterations. Therefore, the decrease of the AE function (or the GAP function) at the beginning of the decoding process is more important than at the end, since the rapid decrease of the uncertainty of the value of variable nodes at the beginning of the decoding might lead to a quick early stop of the decoding process. Thus, scheduling designs at the later stage of decoding may not be used at all. As a result, it is not practical to simply use the average decline rate of $AE_{\boldsymbol{\beta}}(\boldsymbol{c},t)$ (or $GAP_{\boldsymbol{\beta}}(\boldsymbol{c},t)$) in decoding to measure the performance of the corresponding scheduling sequence.

Due to the aforementioned reasons, we define a function $\tau_{AE}(\boldsymbol{c}, \boldsymbol{\beta})$ as a deformed overall decline speed of the function $AE_{\boldsymbol{\beta}}(\boldsymbol{c},t)$ with channel messages with L-density $\boldsymbol{c}=(c_1,\cdots,c_N)$ under a given scheduling sequence $\boldsymbol{\beta}$ . It is defined as follows:

\begin{equation}
\tau_{AE}(\boldsymbol{c}, \boldsymbol{\beta})= \sum_{t=1}^{T}{ AE_{\boldsymbol{\beta}}(\boldsymbol{c},t)},
\end{equation}
where $T$ denotes the maximum number of NMP used in decoding. It is evident that the decrease in $AE_{\boldsymbol{\beta}}(\boldsymbol{c},t)$ at the beginning of decoding will significantly affect $\tau_{AE}(\boldsymbol{c}, \boldsymbol{\beta})$, aligning with actual requirements. Under this definition, we consider that a smaller value of $\tau_{AE}(\boldsymbol{c}, \boldsymbol{\beta})$ corresponds to higher decoding efficiency under the corresponding scheduling sequence $\boldsymbol{\beta}$. We can similarly define a deformed overall decline speed $\tau_{GAP}(\boldsymbol{c}, \boldsymbol{\beta})$ of the function $GAP_{\boldsymbol{\beta}}(\boldsymbol{c},t)$ with channel messages with L-density $\boldsymbol{c}=(c_1,\cdots,c_N)$ under a given scheduling sequence $\boldsymbol{\beta}$ :

\begin{equation}
\tau_{GAP}(\boldsymbol{c},\boldsymbol{\beta})= \sum_{t=1}^{T}{ GAP_{\boldsymbol{\beta}}(\boldsymbol{c},t)}.
\end{equation}

Therefore, our objective is to find the scheduling sequence $\boldsymbol{\beta}$ that minimizes $\tau_{AE}(\boldsymbol{c}, \boldsymbol{\beta})$ (or $\tau_{GAP}(\boldsymbol{c}, \boldsymbol{\beta})$) for a given L-density $\boldsymbol{c}$.

In order to find an optimal scheduling sequence, we propose an optimization algorithm called Successive-Searching BP (SSBP), which is based on the local search algorithm. SSBP starts with an initial sequence and iteratively searches for a scheduling sequence that has a lower value of $\tau_{AE}(\boldsymbol{c}, \boldsymbol{\beta})$ (or $\tau_{GAP}(\boldsymbol{c}, \boldsymbol{\beta})$) than the current sequence by randomly modifying a part of the current sequence. This process continues until no better scheduling sequence can be found.

To reduce the search space and facilitate scheduling, we adopt a CN-centric scheduling policy, which selects a check node in each decoding step and updates all information connected to this node simultaneously, resulting in a check node sequence as the corresponding scheduling sequence. This way, instead of updating $AE_{\boldsymbol{\beta}}(\boldsymbol{c},t)$ (or $GAP_{\boldsymbol{\beta}}(\boldsymbol{c},t)$) every time a message is passed while computing $\tau_{AE}(\boldsymbol{c}, \boldsymbol{\beta})$ (or $\tau_{GAP}(\boldsymbol{c}, \boldsymbol{\beta})$), we update it only after all messages connected to the chosen check node passed. Moreover, we limit the maximum number of decoding iterations to $T$, and the same scheduling sequence is used in each iteration. It is worth mentioning that although we adopt a CN-centric scheduling policy, our approach is also applicable to QC-block-centric scheduling policy and other scheduling policies.

We have investigated the impact of different initial scheduling sequences on the SSBP algorithm. We tested the cases where the initial scheduling sequence was selected by layered scheduling, nested scheduling, and LD scheduling, and the results showed significant improvements in all these cases. 

After constructing the initial scheduling sequence $\boldsymbol{\beta_0}$ and obtaining the L-density $\boldsymbol{c}=(c_1,\cdots,c_N)$ of the channel messages, we compute the corresponding $\tau_{AE}(\boldsymbol{c}, \boldsymbol{\beta_0})$ (or $\tau_{GAP}(\boldsymbol{c}, \boldsymbol{\beta_0})$) value of $\boldsymbol{\beta_0}$, denoted as $\tau_0$. We then select $h$ pairs of check nodes uniformly at random and change the order of these pairs sequentially in $\boldsymbol{\beta_0}$. This process is repeated $b$ times to generate $b$ new sequences from $\boldsymbol{\beta_0}$, and we calculate the corresponding $\tau_{AE}(\boldsymbol{c}, \boldsymbol{\beta})$ (or $\tau_{GAP}(\boldsymbol{c}, \boldsymbol{\beta})$) value for each new sequence. The scheduling sequence with the minimum $\tau_{AE}$ (or $\tau_{GAP}$) among the $\boldsymbol{\beta_0}$ and $b$ new sequences will replace the original $\boldsymbol{\beta_0}$ to become the new starting sequence in the next round of SSBP. We continue this process from the new starting sequence to find better scheduling sequences. If none of the new sequences obtained are better than the original scheduling sequence, we consider the sequence update to have failed. The algorithm terminates when the sequence update fails $S$ consecutive times.

The details of SSBP are shown in Algorithm 1. In line 6 of Algorithm 1, the exchange function obtains the descendants of the sequence $\boldsymbol{\beta_0}$ in each round of the SSBP algorithm by randomly modifying
a part of $\boldsymbol{\beta_0}$. The exchange function is described in detail in Algorithm 2, where line 1 retrieves the scheduling sequence of the first iteration, and lines 3-8 randomly exchange the scheduling order $h$ times.

\begin{algorithm}
	\renewcommand{\algorithmicrequire}{\textbf{Input:}}
	\renewcommand{\algorithmicensure}{\textbf{Output:}}
	\caption{Successive-Searching BP (SSBP) algorithm}
	\label{alg:1}
	\begin{algorithmic}[1]
		\REQUIRE the parity check metric H of the LDPC code, the number of check nodes $M$, the number of max iteration $T$, the channel messages with the L-density $\boldsymbol{c}=(c_1,\cdots,c_N)$
		\ENSURE scheduling sequence $\boldsymbol{\beta}$
		\STATE Initial the scheduling sequence $\boldsymbol{\beta_0}$.
		\STATE $\tau_{0} \gets \tau_{AE}(\boldsymbol{c}, \boldsymbol{\beta_0})$ (or $\tau_{0} \gets \tau_{GAP}(\boldsymbol{c}, \boldsymbol{\beta_0})$)
		\STATE $s\gets0$
		\WHILE{$s\leq S$}
		\FOR{$i=1:b$}
		\STATE  $\boldsymbol{\beta_i}\gets$Exchange($\boldsymbol{\beta_0}$, $M$, $T$)
		\STATE $\tau_{i} \gets \tau_{AE}(\boldsymbol{c}, \boldsymbol{\beta_i})$ (or $\tau_{i} \gets \tau_{GAP}(\boldsymbol{c}, \boldsymbol{\beta_i})$)
		\ENDFOR
		\STATE $(\tau',\boldsymbol{\beta'})\gets min(\tau_{i},\boldsymbol{\beta_{i}})$($min$ takes over $\tau_i$, $i=\left\{0,1,\cdots,b\right\}$)
		\IF{$\boldsymbol{\beta'}==\boldsymbol{\beta_0}$}
		\STATE $s\gets s+1$
		\STATE \textbf{continue}
		\ENDIF
		\STATE $\tau_0\gets\tau'$
		\STATE $\boldsymbol{\beta_0}\gets\boldsymbol{\beta'}$
		\ENDWHILE
		\STATE $\boldsymbol{\beta}\gets\boldsymbol{\beta_0}$
		\STATE \textbf{return} $\boldsymbol{\beta}$
	\end{algorithmic}  
\end{algorithm}

\begin{algorithm}
	\renewcommand{\algorithmicrequire}{\textbf{Input:}}
	\renewcommand{\algorithmicensure}{\textbf{Output:}}
	\caption{Exchange algorithm}
	\label{alg:1}
	\begin{algorithmic}[1]
		\REQUIRE current scheduling sequence $\boldsymbol{\beta_0}$, the number of check nodes $M$ and maximum iteration number $T$
		\ENSURE new scheduling sequence $\boldsymbol{\beta_i}$
		\STATE $\boldsymbol{\gamma_0}\gets\boldsymbol{\beta_0}\left[1:M\right]$ (Pick out bits 1 to $M$ of the sequence $\beta_0$ to form the sequence $\gamma_0$ of length $M$)
		\FOR{$i=1:h$}
		\STATE Choose $j,k\in \left\{1,\cdots,M\right\}$ uniformly at random
		\STATE $\boldsymbol{\gamma_0}_{temp} \gets \boldsymbol{\gamma_0} \left[j\right]$
		\STATE $\boldsymbol{\gamma_0}\left[j\right] \gets \boldsymbol{\gamma_0}\left[k\right]$
		\STATE $\boldsymbol{\gamma_0}\left[k\right] \gets \boldsymbol{\gamma_0}_{temp}$
		\ENDFOR
		\STATE Copy the sequence $\boldsymbol{\gamma_0}$ $T$ times to get new scheduling sequence $\boldsymbol{\beta_i}$ of length $MT$
		\STATE $\textbf{return}$ $\boldsymbol{\beta_i}$
	\end{algorithmic}  
\end{algorithm}

\subsection{Average-case computational complexity analysis}
Following the analysis of the average-case computational complexity of the local search algorithm in \cite{aarts2003local}, an upper bound of the average number of rounds SSBP required is stated in the next proposition.

\begin{prop}
The average number of rounds SSBP required is less than $SM^{2h+1}$.

\begin{proof}
Let $G = (V, E)$ be an undirected graph, where $V$ is composed of all possible scheduling sequences. Then $|V|=M!$, since the number of check nodes is $M$ and the same scheduling sequence is adopted in each iteration. If scheduling sequence $i$ can change into scheduling sequence $j$ in one Exchange algorithm, then $\{i,j\} \in E$. It is clear that $G$ is a regular graph. Denote the degree of $G$ by $d_G$. Then

\begin{equation}
d_G \leq \binom{M}{2}^h < \frac{1}{2^h}M^{2h},
\end{equation}
where the first inequality follows from the fact that there are $\binom{M}{2}$ options when selecting one pair of check nodes for exchange, and $h$ pairs of check nodes are needed to be selected.

The SSBP algorithm starts from an initial vertex in $G$ and continues to move to a better neighborhood of the current vertex. This process can correspond to a path on $G$, where the value of the function $\tau_{AE}$ (or $\tau_{GAP}$) will continuously decrease along the vertices on $P$. Therefore, in the average case, if no path in $G$ of length $M^{2h+1}$ has the property of decreasing the $\tau_{AE}$ (or $\tau_{GAP}$) function value, Proposition 1 is proved, since SSBP may take $S$ rounds to move to the next vertex.

Denote the number of paths of length $M^{2h+1}$ in the graph $G$ as $|\mathcal{P}|$. Since these paths can start at any point in $G$,

\begin{equation}
|\mathcal{P}| \leq M!(d_G)^{M^{2h+1}} < \frac{M!}{2^h}M^{2hM^{2h+1}}.
\end{equation}

For any fixed path of length $M^{2h+1}$, there are $M^{2h+1}!$ possible arrangements of the value of the function $\tau_{AE}$ (or $\tau_{GAP}$), and each possible arrangement is equally likely to occur. Therefore, the number of paths of length $M^{2h+1}$ in the graph $G$ with decreasing $\tau_{AE}$ (or $\tau_{GAP}$) value is less than

\begin{equation}
\frac{M!}{2^h}\frac{M^{2hM^{2h+1}}}{M^{2h+1}!} < 1,
\end{equation}
where the last inequality follows from Stirling's approximation.
\end{proof}
\end{prop}

\begin{figure*}
\centering

\includegraphics[width=18cm]{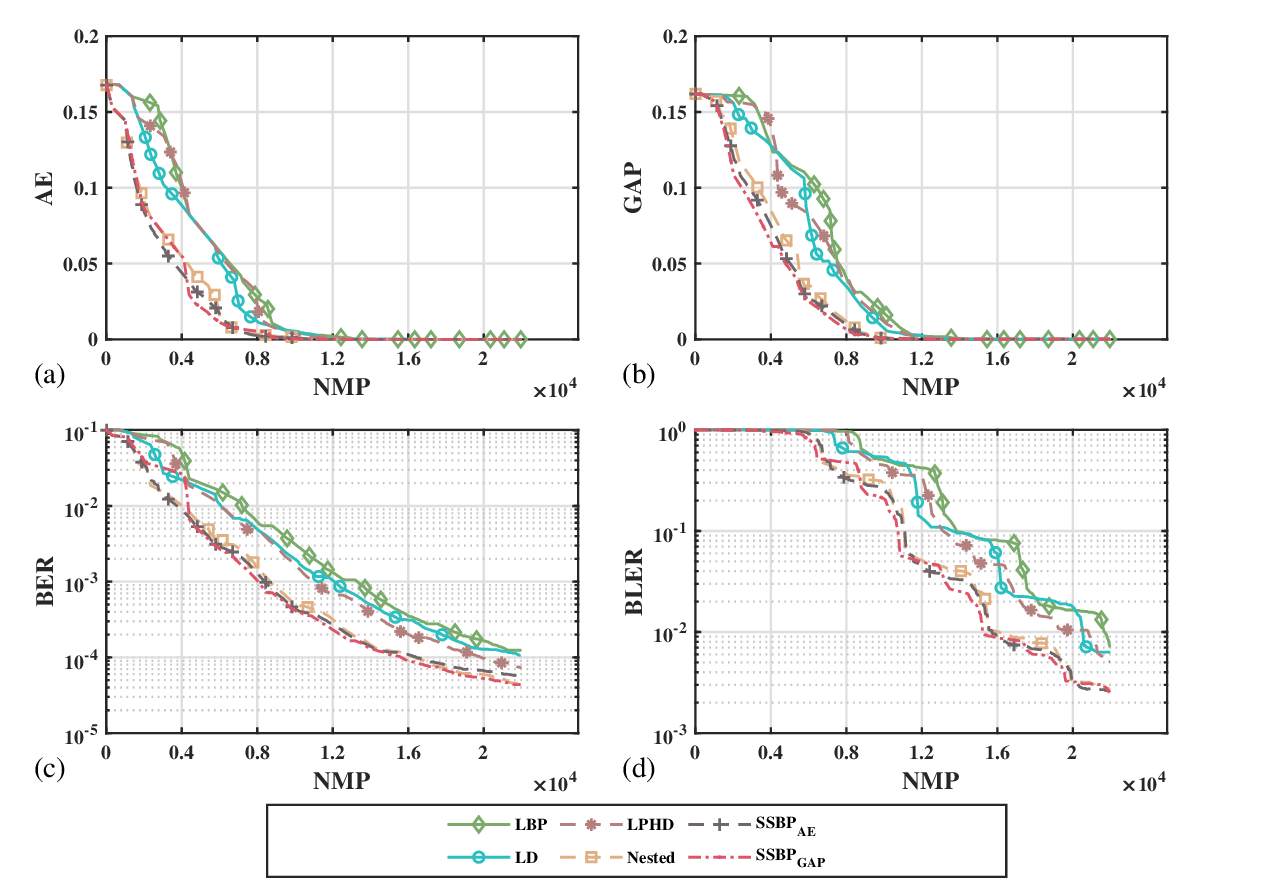}
\caption{Average entropy (AE), GAP, BER, and BLER vs. NMP for different scheduling policies in (512, 384) BG1 graph, where the number of iterations is fixed to be 5. SSBP has the fastest descent speed and better error-rate performance.}
\label{fig}
\end{figure*}

\section{Simulation Results}
In this section, we apply the SSBP algorithm to LDPC codes constructed following 5G NR. We test the coding scheme (1024, 768), (512, 384) for base graph 1 (BG1), and (512, 256), (256, 64) for base graph 2 (BG2). By calculating the average NMP under different scheduling policies, we compare the SSBP algorithm for the AE case and SSBP for the GAP case with layered BP (LBP) \cite{hocevar2004reduced}, least-punctured and highest-degree (LPHD) scheduling policy \cite{wang2020two}, lowest-degree (LD) scheduling policy \cite{frenzel2019static}, and nested scheduling policy \cite{jang2022design}. The scheduling policy of LBP follows the order of rows in the parity-check matrix.
% \begin{table*}
% \centering
% \caption{AVERAGE NMP FOR SEVERAL SCHEDULING POLICIES \protect\\SSBP can reduce NMP by $20\%$ compared with LBP in some cases.}
% \label{table}
% \setlength{\tabcolsep}{3pt}
% \begin{tabular}{|c|c|c|c|c|c|c|c|c|}
% \hline
%  BG type&
%  N &
%  K &
%  Eb/N0 &
%  LBP & 
%  LPHD &
%  LD &
%  SSBP(entropy)&
%  SSBP(GAP)\\
% \hline
% BG1 &
% 512 &
% 384 &
% 6.0 &
% 5557.19&
% 5188.05&
% 5115.01&
% 4419.57&
% 4480.69\\
% \hline
% BG1 &
% 1024 &
% 768 &
% 5.8 &
% 12342.72 &
% 12185.63 &
% 12200.71 &
% 10637.25 &
% 10785.75\\
% \hline
% BG2 &
% 512 &
% 256 &
% 3.0 &
% 5948.16 &
% 5525.61 &
% 5773.70 &
% 5256.54 &
% 5297.90\\
% \hline
% BG2 &
% 256 &
% 64 &
% 1.5 &
% 2051.45 &
% 1978.30 &
% 1713.74 &
% 1681.65 &
% 1732.80\\
% \hline
% \end{tabular}
% \label{tab1}
% \end{table*}

\begin{table*}[]
\centering
\caption{AVERAGE NMP FOR SEVERAL SCHEDULING POLICIES IN BG1. \protect\\ SSBP can reduce NMP by more than $20\%$ compared with LBP.}
\label{table3}
\setlength{\tabcolsep}{3pt}
\begin{threeparttable}
\begin{tabular}{|l|lll|lll|}
\hline
                      & \multicolumn{3}{c|}{{ BG1 N=512 R=0.75 Eb/N0=6.0}} & \multicolumn{3}{c|}{{ BG1 N=1024 R=0.75 Eb/N0=5.8}} \\ \hline
Scheduling policies   & \multicolumn{1}{l|}{Average NMP}           & \multicolumn{1}{l|}{Reduction ratio\tnote{1}} & BLER ($\times 10^{-3})$\tnote{2}           & \multicolumn{1}{l|}{Average NMP}           & \multicolumn{1}{l|}{Reduction ratio}  & BLER ($\times 10^{-3}$)         \\ \hline
LBP \cite{hocevar2004reduced}                  & \multicolumn{1}{l|}{11114.38}               & \multicolumn{1}{l|}{0} &            6.98            & \multicolumn{1}{l|}{24685.44}               & \multicolumn{1}{l|}{0}&            5.21          \\ \hline
LPHD  \cite{wang2020two}                & \multicolumn{1}{l|}{10376.10}               & \multicolumn{1}{l|}{6.64\%}          &  5.09        & \multicolumn{1}{l|}{24371.26}               & \multicolumn{1}{l|}{1.27\%}   &      2.94           \\ \hline
LD  \cite{frenzel2019static}                  & \multicolumn{1}{l|}{10230.02}               & \multicolumn{1}{l|}{7.96\%}     &     6.32          & \multicolumn{1}{l|}{24401.42}               &\multicolumn{1}{l|}{1.15\%}   &           3.69      \\ \hline
Nested \cite{jang2022design}                 & \multicolumn{1}{l|}{8185.27}               & \multicolumn{1}{l|}{26.35\%}     &     2.70          & \multicolumn{1}{l|}{19949.60}               &\multicolumn{1}{l|}{19.18\%}   &           1.15      \\ \hline
SSBP(AE) & \multicolumn{1}{l|}{8064.28}               & \multicolumn{1}{l|}{27.44\%}&         2.60          & \multicolumn{1}{l|}{19338.32}               & \multicolumn{1}{l|}{21.66\%} &        0.96          \\ \hline
SSBP(GAP)             & \multicolumn{1}{l|}{7873.02}               & \multicolumn{1}{l|}{29.16\%} &     2.51             & \multicolumn{1}{l|}{19900.32}               & \multicolumn{1}{l|}{19.38\% } &    1.15             \\ \hline
\end{tabular}
 \begin{tablenotes}
        \footnotesize
        \item[1] The reduction ratio shows the reduction of the average NMP under the corresponding scheduling policies compared with the average NMP under LBP.
        \item[2] The BLER$(\times 10^{-3})$ shows the block error rate under the corresponding scheduling policies, where the max number of iterations is set to be 5, and the results are displayed after multiplying by $10^{3}$.
      \end{tablenotes}
  \end{threeparttable}
\end{table*}

\begin{table*}[]
\centering
\caption{AVERAGE NMP FOR SEVERAL SCHEDULING POLICIES IN BG2.}
\label{table4}
\setlength{\tabcolsep}{3pt}
\begin{tabular}{|l|lll|lll|}
\hline
                      & \multicolumn{3}{c|}{{ BG2 N=512 R=0.50 Eb/N0=3.0}} & \multicolumn{3}{c|}{{ BG2 N=256 R=0.25 Eb/N0=1.5}} \\ \hline
Scheduling policies   & \multicolumn{1}{l|}{Average NMP}           & \multicolumn{1}{l|}{Reduction ratio} & BLER ($\times 10^{-3}$)           & \multicolumn{1}{l|}{Average NMP}           & \multicolumn{1}{l|}{Reduction ratio}  & BLER ($\times 10^{-4}$)         \\ \hline
LBP                   & \multicolumn{1}{l|}{11896.32}               & \multicolumn{1}{l|}{0} &       7.84                 & \multicolumn{1}{l|}{4102.90}               & \multicolumn{1}{l|}{0}&            3.88          \\ \hline
LPHD                  & \multicolumn{1}{l|}{11051.22}               & \multicolumn{1}{l|}{7.10\%}          &    5.64      & \multicolumn{1}{l|}{3956.60}               & \multicolumn{1}{l|}{3.56\%}   &    2.30             \\ \hline
LD                    & \multicolumn{1}{l|}{11547.40}               & \multicolumn{1}{l|}{2.93\%}     &   8.57            & \multicolumn{1}{l|}{3427.48}               &\multicolumn{1}{l|}{16.46\%}   &        0.57         \\ \hline
Nested                    & \multicolumn{1}{l|}{10060.50}               & \multicolumn{1}{l|}{15.43\%}     &   5.82            & \multicolumn{1}{l|}{3200.71}               &\multicolumn{1}{l|}{21.99\%}   &        0.47         \\ \hline
SSBP(AE) & \multicolumn{1}{l|}{9620.89}               & \multicolumn{1}{l|}{19.12\%}&    4.15               & \multicolumn{1}{l|}{3063.97}               & \multicolumn{1}{l|}{25.32\%} &         0.40         \\ \hline
SSBP(GAP)             & \multicolumn{1}{l|}{9834.34}               & \multicolumn{1}{l|}{17.33\%} &         4.94         & \multicolumn{1}{l|}{3235.10}               & \multicolumn{1}{l|}{21.15\% } &          0.83       \\ \hline
\end{tabular}
\end{table*}

To accommodate high throughput application scenarios, we limit the maximum number of decoding iterations to 5 in all our experiments. For the hyperparameters in SSBP, we set $b$, the number of searches at one time, to be 100, and $S$, the stop criterion, is set to 10. The selection of parameters $b$ and $S$ will have an impact on the algorithm's runtime and effectiveness. Increasing the values of $b$ and $S$ to some extent allows the algorithm to find better scheduling sequences but also prolongs the runtime. Experiments show that the chosen values of $b$ and $S$ strike a good balance between the algorithm's runtime and effectiveness. Regarding the number of check node pairs for change $h$, experiments show that continuous optimization of the scheduling sequence in the algorithm is only possible when $h$ is relatively small, such as $\lfloor 0.5\% \times M \rfloor$ in our simulation experiments. The possible reason for this is that most sequences in the entire sequence space are bad for decoding. Therefore, a new scheduling sequence obtained from the original sequence by changing a large part of the original sequence is likely to be a "bad" scheduling sequence, which is detrimental to finding a better scheduling sequence than the current one. The average NMP of these scheduling policies on several 5G NR graphs is shown in TABLE $\rm{\uppercase\expandafter{\romannumeral1}}$ and TABLE $\rm{\uppercase\expandafter{\romannumeral2}}$, where $N$ is the length of the codeword, and $R$ is the code rate.

As shown in TABLE $\rm{\uppercase\expandafter{\romannumeral1}}$ and TABLE $\rm{\uppercase\expandafter{\romannumeral2}}$, the SSBP scheduling policy can significantly reduce the average NMP required for the decoding process, both for the AE case and for the GAP case. In the (512, 384) BG1, the SSBP for the AE case can reduce NMP by almost $30\%$ compared to the LBP algorithm. Additionally, as previously mentioned, SSBP is an offline BP strategy, which means that we can determine the decoding order through Algorithm 1 before the formal decoding process. Therefore, SSBP can achieve remarkable decoding complexity gains with almost no additional scheduling complexity added to the formal decoding process, as compared to LBP.

To gain a clearer understanding of the effect of different scheduling policies on decoding, we present in Figure 1 the changes in AE, GAP, bit error rate (BER), and block error rate (BLER) with NMP changing when decoding under different scheduling policies for the (512, 384) BG1, where the number of iterations is fixed at 5. We observe that as NMP changes, AE and GAP change similarly to the BER and BLER, with LBP always being the slowest to change, while SSBP for the AE case and SSBP for the GAP case always have the fastest-changing speed. This demonstrates that AE and GAP can effectively reflect the performance of the decoding process, and scheduling sequences obtained by SSBP have better performance throughout the decoding process. Thus, SSBP can speed up the decoding process and reduce the complexity of decoding. Furthermore, SSBP can achieve lower BER and BLER with a limited maximum number of iterations compared to the other scheduling policies. For instance, for the (512, 384) BG1, the BER for LBP after 5 iterations is $1.24\times 10^{-4}$, while the BER for SSBP for the AE case and the GAP case are only $5.66\times 10^{-5}$ and $4.34\times 10^{-5}$, respectively, which are more than $50\%$ lower than LBP.

\section{Conclusion}
In this study, we presented an efficient scheduling policy for LDPC codes. Firstly, we proposed a model to evaluate the complexity and performance of the decoding process using three metrics, namely NMP, AE, and GAP. Based on this model, we designed the SSBP offline scheduling algorithm and analyzed the upper bound of the average-case computational complexity required by SSBP. Our simulation results demonstrate that SSBP can significantly improve the convergence speed and achieve better error-rate performance without introducing extra scheduling complexity in the decoding process. It should be noted that other optimization methods, such as machine learning and ant colony optimization, can also be applied to our proposed model to further improve the efficiency of offline scheduling policies. Moreover, better optimization algorithms for finding well-performing scheduling sequences and the theoretically optimal value of scheduling sequences remain to be further studied.

\section*{Acknowledgment}
I would like to express my gratitude to Longlong Li for his valuable discussion on this study.

\bibliographystyle{unsrt}
\bibliography{reference}

\end{document}